\documentclass[cameraready]{Interspeech}
\title{
Improving Text-to-Audio Instruction Following via Fine-Grained Feedback from Audio-Aware Large Language Models
}
\newcommand\blfootnote[1]{%
  \begingroup
  \renewcommand\thefootnote{}\footnote{#1}%
  \addtocounter{footnote}{-1}%
  \endgroup
}

\author[affiliation={1,2}]{Chun-Yi}{Kuan\texorpdfstring{\,${}^{\clubsuit}$}{}}

\author[affiliation={2}]{Siwon}{Kim}
\author[affiliation={2}]{Byeonggeun}{Kim}
\author[affiliation={2}]{Suyoun}{Kim}
\author[affiliation={2}]{Bo-Ru}{Lu}
\author[affiliation={2}]{Qingming}{Tang}
\author[affiliation={2}]{Ankur}{Gandhe}
\author[affiliation={1}]{Hung-yi}{Lee}
\author[affiliation={2}]{Chieh-Chi}{Kao}
\author[affiliation={2}]{Chao}{Wang}

\address{
    $^1$National Taiwan University, Taiwan, $^2$Amazon, USA
}

\email{chunyi.kuan.tw@gmail.com}
\keywords{text-to-audio generation, audio-aware large language models, LLM-based evaluation, instruction following}

\usepackage{comment}
\usepackage{amsmath,graphicx,hyperref}
\usepackage{cite}
\usepackage{multirow}
\usepackage{url} 
\usepackage{amsmath,amssymb,amsfonts}
\usepackage{algorithmic}
\usepackage{adjustbox}
\usepackage{textcomp}
\usepackage{booktabs}
\usepackage{cleveref}

\usepackage[table]{xcolor}
\definecolor{lightgray}{RGB}{240,240,240}
\definecolor{lightblue}{RGB}{220,235,255}
\definecolor{lightyellow}{RGB}{255,250,205}

\definecolor{lightgreen}{HTML}{EAF2E3}
\definecolor{lightorange}{HTML}{FFF3E6}

\begin{document}

\maketitle
% $^{\heartsuit}$   % 愛心 ♥
% $^{\clubsuit}$    % 梅花 ♣
% $^{\spadesuit}$   % 黑桃 ♠
% $^{\diamondsuit}$ % 方塊 ♦

% \blfootnote{$^{*}$Work done during internship at Amazon.}
% \blfootnote{$^{\clubsuit}$Work done during internship at Amazon.}
\blfootnote{\textsuperscript{\ensuremath{\clubsuit}}\,Work done during internship at Amazon.}

% $^{\heartsuit}$
% $^{\dagger}$
\begin{abstract}

Recent text-to-audio models generate high-quality audio, but often fail to follow instructions involving multiple sound events and temporal ordering. 
This gap arises because existing evaluation and training signals mainly emphasize global similarity or perceptual quality, with limited supervision on instruction-level correctness. 
We propose an instruction-level framework that uses audio-aware large language models (ALLMs) as fine-grained judges to verify target event presence and temporal relations in generated audio. 
After validating ALLM judgments on benchmarks and through human verification, we use their feedback to construct preference pairs for direct preference optimization.
We further introduce S3Bench, a narrative benchmark for evaluating multi-event temporal instruction following. 
Experiments show that our method improves event completeness, temporal ordering, and joint instruction-following accuracy across existing benchmarks and S3Bench, while maintaining audio quality.
\end{abstract}

\section{Introduction}
\begin{figure*}[ht]
    \centering
    \includegraphics[width=0.85\textwidth]{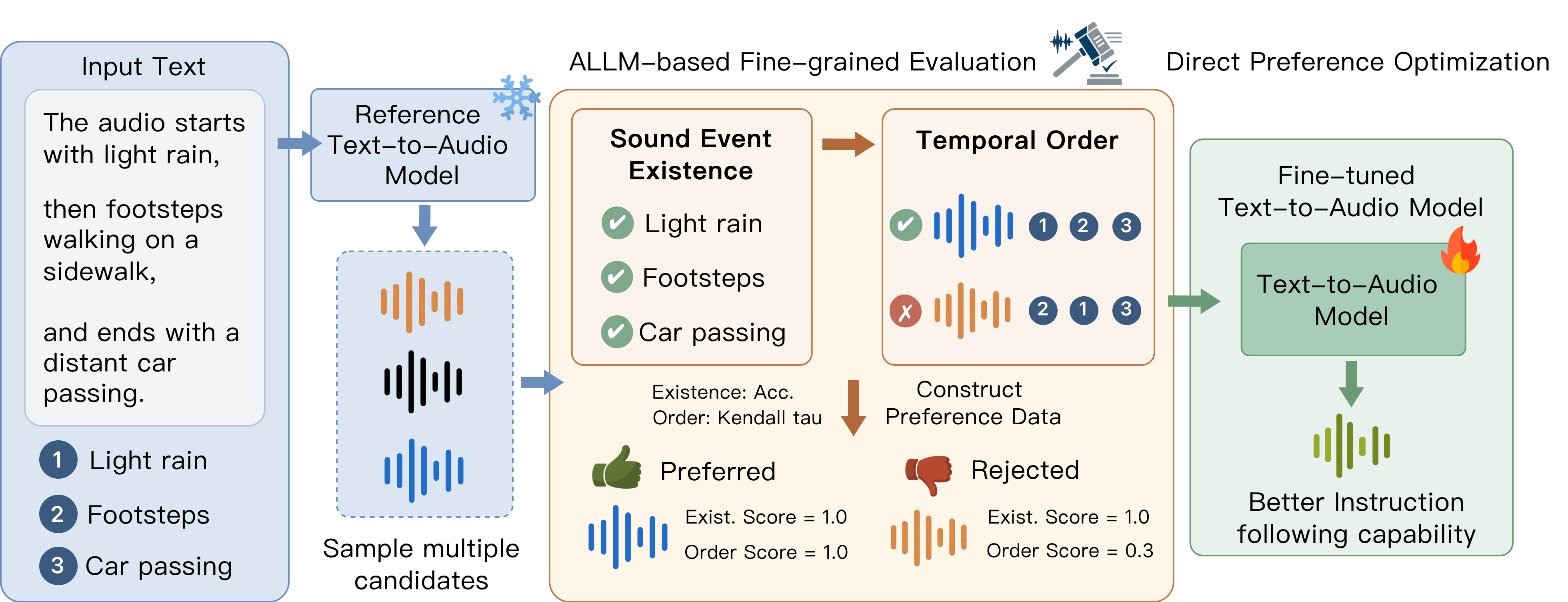} 
    \caption{
    % Overview of our framework.
    % Fine-grained ALLM evaluation of generated audios is used to construct preference data for DPO-based text-to-audio model fine-tuning.
    % Overview of the proposed framework. 
    % Fine-grained ALLM judgments are used to construct preference data for direct preference optimization training.
    Overview of the proposed ALLM-Judged Preference Optimization framework. Given a text instruction, a TTA model generates multiple candidate audio clips, which are then evaluated by an audio-aware large language model for sound event existence and temporal ordering. The resulting fine-grained judgments are converted into preference pairs for direct preference optimization.
    }
    \label{fig:overview}
\end{figure*}

Recent progress in text-to-audio (TTA) generation\cite{kreuk2022audiogen,agostinelli2023musiclm,liu2023audioldm,ghosal2023tango,yang2023diffsound,kuan2023towards,copet2023simple,borsos2023audiolm,borsos2023soundstorm,huang2023make2,majumder2024tango,hung2024tangofluxsuperfastfaithful,hai2024ezaudio,audioldm2-2024taslp,evans2025stable,leeetta,huangimpact,yanggenerative} 
has led to substantial improvements in perceptual audio quality, enabling models to produce increasingly realistic and natural-sounding audio from textual descriptions.
Modern TTA systems, based on diffusion or flow-matching frameworks, achieve strong performance on widely used evaluation metrics such as FAD~\cite{kilgour2019fad}, IS~\cite{salimans2016improved}, and CLAPScore~\cite{elizalde2023clap, wu2023large}, indicating high overall audio fidelity and semantic alignment at a coarse level.
% on widely used evaluation metrics
% on widely adopted benchmarks
However, beyond perceptual realism, a core requirement of practical TTA systems is the ability to faithfully follow fine-grained textual instructions, particularly when the input specifies multiple sound events and their temporal relationships. 
For example, users might ask TTA systems to generate audio based on the text: ``It starts with gentle rainfall, then someone walks through the scene. After that, a wooden door slowly opens, and the clip finishes with cars passing by in the distance.''
Recent studies
~\cite{
majumder2024tango, 
hung2024tangofluxsuperfastfaithful, 
he2025ritta, he2026aurelius} 
have shown that existing TTA models often struggle to generate all target sound events within a single audio clip and to produce them in the correct temporal order, even when such constraints are explicitly stated in the text.
We also conduct a systematic evaluation of current TTA systems on our proposed evaluation sets, revealing room for improvement in multi-event generation and temporal order control.

A key reason for this limitation is the lack of scalable instruction-level feedback for text-to-audio generation.
Most current evaluation metrics and training objectives focus on global audio-text similarity~\cite{elizalde2023clap, wu2023large} or perceptual quality, while providing little supervision on instruction-level correctness, such as event completeness and temporal ordering.
More fine-grained feedback that measures whether each target sound event is generated and whether the specified temporal order is satisfied would provide a clearer and more diagnostic assessment of a TTA system's instruction-following capability.
However, several prior studies~\cite{ghoshcompa, wang2025t2a, kuan2026aqascore} have reported that widely used metrics such as CLAP~\cite{elizalde2023clap, wu2023large} struggle to capture temporal relationships between sound events.
As a result, a generated clip may receive a high similarity score while still missing secondary events or reversing the intended event sequence, creating a mismatch between what current objectives reward and what users expect from instruction-following TTA systems.
NegEval-Audio~\cite{kuan2026negaudio} further shows that CLAP-style embeddings poorly distinguish between captions describing present and absent sound events.

To provide such instruction-level feedback, we turn to recent audio-aware large language models (ALLMs)~\cite{achiam2023gpt, gong2023joint, wang2023blsp, fathullah2023towards, kuan2024speech, chang2024speechprompt, wang2024blsp, goel2025audio, liu2025voxtral, xu2025qwen2, kuan2025teaching, kuan2025alignment, abouelenin2025phi, comanici2025gemini, arora2025landscape, ghoshaudio}, which have demonstrated strong audio understanding and reasoning abilities on several audio benchmarks~\cite{kuan2024understanding, kuan2025can, huang2024dynamic, huang2024dynamic2, sakshimmau, kumar2025mmau, ma2025mmar, chang2025game, kuan2026}, including MMAU~\cite{sakshimmau}.
These capabilities suggest a promising opportunity: instead of relying only on global similarity scores, ALLMs may be used to provide structured feedback on whether generated audio satisfies fine-grained textual instructions.
In this work, we explore whether ALLM-based verification can serve not only as an evaluation tool, but also as a training signal for improving TTA instruction following.
This also bridges a gap between audio understanding and audio generation: while ALLMs exhibit strong semantic and temporal reasoning abilities, their judgments have rarely been used as training signals for improving TTA generation.

While ALLMs offer a promising source of instruction-level feedback, several recent studies~\cite{majumder2024tango, wang2025t2a, liao2024baton} have also begun to address multi-event generation and temporal ordering in TTA through preference data construction, heuristic perturbations, or human annotations.
% Several recent studies~\cite{majumder2024tango, wang2025t2a, liao2024baton} have begun to address multi-event generation and temporal ordering in TTA through preference data construction, heuristic perturbations, or human annotations.
For example, Tango2~\cite{majumder2024tango} constructs Audio-Alpaca by filtering samples with CLAPScore and generating negative examples with reversed event orders.
T2A-Feedback~\cite{wang2025t2a} uses a cascade pipeline that decomposes instructions, separates generated audio into event-level segments, and computes event-wise CLAP similarities to form preference data.
Baton~\cite{liao2024baton} collects human preference annotations focusing on sound-event completeness and temporal relationships.
While these approaches represent important steps toward fine-grained supervision, they often rely on CLAP-based matching, handcrafted perturbations, complex heuristic pipelines, or costly human feedback collection.
In contrast, we use ALLMs as scalable instruction-level judges that directly reason over generated audio and the input text, producing explicit judgments on event existence and temporal order.

Our framework proceeds in three steps.
First, we empirically verify that recent ALLMs can assess sound event existence and temporal ordering on established audio understanding benchmarks with ground-truth annotations.
% To further examine their reliability in practical TTA settings, we conduct a human verification study on a subset of TTA-generated audio and measure the agreement between ALLM judgments and human judgments.
Second, we use the ALLM judgments to score multiple candidate generations for each instruction.
To further examine their reliability in practical TTA settings, we conduct a human verification study on a subset of TTA-generated audio and measure the agreement between ALLM and human judgments. 
% We then use the ALLM feedback to score multiple candidate generations for each instruction.
Samples that satisfy both event presence and temporal order are treated as preferred, while samples with missing events or incorrect ordering are treated as rejected.
These structured preference pairs are then used to optimize a TTA model with direct preference optimization.
Third, to evaluate TTA models under more demanding multi-event and temporal-order conditions, we introduce the Sound Scene Story Benchmark (S3Bench), a narrative benchmark containing instructions with multiple sound events and explicit temporal progression.
Experiments show that our method improves event completeness, temporal ordering, and joint instruction-following accuracy across existing benchmarks and S3Bench, while maintaining competitive audio quality.

Our main contributions are summarized as follows.
First, we formulate fine-grained instruction following in TTA generation around two explicit criteria: sound event existence and temporal ordering, which capture instruction completeness and faithfulness beyond global audio-text similarity.
Second, we show that recent ALLMs can serve as effective instruction-level judges for these criteria using established audio understanding benchmarks and a human verification study on TTA-generated audio.
Third, we propose ALLM-Judged Preference Optimization, which converts structured ALLM judgments into preference pairs for DPO-based TTA training.
Finally, we introduce the Sound Scene Story Benchmark, a narrative multi-event benchmark for evaluating temporal instruction following, and show that our method improves event completeness, temporal ordering, and joint instruction-following accuracy while maintaining competitive audio quality.
% Generated samples, side-by-side comparisons with baselines, and additional evaluation results are available on our \href{https://kuan2jiu99.github.io/allm-feedback-tta}{demo website}.
% Additional qualitative examples, comparative analyses with baseline systems, and extended evaluation results are provided on the \href{https://kuan2jiu99.github.io/allm-feedback-tta}{project page}.
% A \href{https://kuan2jiu99.github.io/allm-feedback-tta}{project page} provides additional qualitative examples, comparative analyses with baseline systems, and extended evaluation results from an independent public reproduction based on publicly available models, datasets, and open-source components.
% A \href{https://kuan2jiu99.github.io/allm-feedback-tta}{project page} provides additional qualitative examples, comparative analyses with baseline systems, and extended evaluation results.
A \href{https://kuan2jiu99.github.io/allm-feedback-tta}{project page}\footnote{\scriptsize
\url{https://kuan2jiu99.github.io/allm-feedback-tta}}
provides additional qualitative examples, comparative analyses with baseline
systems, and extended evaluation results.

\section{Related Work}

\subsection{Preference Training for Text-to-Audio Generation}

Preference-based training has recently been adopted to improve text-to-audio (TTA) generation by learning from relative comparisons rather than absolute scores~\cite{majumder2024tango,hung2024tangofluxsuperfastfaithful,wang2025t2a}.
Existing methods mainly construct preference data using similarity-based ranking, heuristic pipelines, or human annotations.
Similarity-based methods, such as Tango2~\cite{majumder2024tango} and TangoFlux-CRPO~\cite{hung2024tangofluxsuperfastfaithful}, rank or filter generated samples using CLAPScore.
Although Tango2 additionally constructs order-related negative examples through prompt perturbation, these methods still rely primarily on global similarity scores rather than direct verification of instruction-level correctness, such as event completeness or temporal order.
T2A-Feedback~\cite{wang2025t2a} provides finer-grained supervision by decomposing instructions, separating generated audio, and computing event-wise CLAP similarities, but relies on a complex cascade pipeline and similarity-based matching.
Baton~\cite{liao2024baton} collects human preferences focusing on sound-event completeness and temporal relationships, offering high-quality supervision at the cost of substantial annotation effort.
In contrast, we use audio-aware large language models as scalable instruction-level judges that directly verify event presence and temporal order, and convert these structured judgments into preference data for TTA training.

\subsection{ALLMs for Audio Understanding and Evaluation}

Recent progress in audio-aware large language models~\cite{achiam2023gpt, gong2023joint, wang2023blsp, fathullah2023towards, kuan2024speech, chang2024speechprompt, wang2024blsp, goel2025audio, liu2025voxtral, xu2025qwen2, kuan2025teaching, kuan2025alignment, abouelenin2025phi, comanici2025gemini, arora2025landscape, ghoshaudio} (ALLMs) has enabled strong audio understanding and reasoning across a wide range of tasks.
Models such as Qwen2.5-Omni~\cite{xu2025qwen2} demonstrate strong performance on a range of audio understanding benchmarks~\cite{kuan2024understanding, kuan2025can, huang2024dynamic, huang2024dynamic2, sakshimmau, kumar2025mmau, ma2025mmar, chang2025game, kuan2026, kuan2026walking}.
% Prior studies~\cite{chenaudio, manakul2025audiojudge, chiang2025audiojudge, huang2025instructttseval, wang2025audioeval} have explored the use of ALLMs as judges for speech and audio evaluation, including descriptive speech quality assessment~\cite{chenaudio}, paralinguistic analysis~\cite{manakul2025audiojudge}, speaking style evaluation~\cite{chiang2025audiojudge}, and instruction-following~\cite{huang2025instructttseval} assessment in text-to-speech systems. 
Prior studies~\cite{chenaudio, manakul2025audiojudge, chiang2025audiojudge, huang2025instructttseval, wang2025audioeval} have explored the use of ALLMs as judges for speech and audio evaluation, including descriptive speech quality assessment~\cite{chenaudio}, paralinguistic analysis~\cite{manakul2025audiojudge, chiang2025audiojudge}, and instruction-following assessment in text-to-speech systems~\cite{huang2025instructttseval}. 
More recently, AudioEval~\cite{wang2025audioeval} proposed a trained ALLM to predict multiple perceptual dimensions of TTA generation, such as production quality, demonstrating the potential of learned evaluators for assessing generative audio models.
Despite these advances, existing studies primarily focus on perceptual quality, speaking styles, or high-level attributes of generated audio. 
% Fine-grained evaluation of instruction-level sound event correctness, such as whether all specified sound events are present and whether they follow the required temporal order, has received relatively limited attention. 
% Moreover, prior work typically treats ALLMs as post-hoc evaluators, without leveraging their judgments to guide TTA model training.
% In contrast, we investigate the use of ALLMs as fine-grained judges for multi-event TTA generation, explicitly evaluating sound event existence and temporal ordering, and further incorporate this feedback into preference-based training to improve instruction-following behavior.
Fine-grained evaluation of instruction-level sound event correctness has received relatively limited attention in existing TTA research. 
In particular, it remains underexplored whether generated audio contains all specified sound events and whether these events occur in the required temporal order.
Moreover, prior work has largely treated ALLMs as post-hoc evaluators, without leveraging their judgments to guide TTA model training.
% In contrast, we investigate the use of ALLMs as fine-grained judges for multi-event TTA generation, explicitly assessing event presence and temporal order, and incorporate their feedback into preference-based training to improve instruction-following behavior.
In contrast, we investigate the use of ALLMs as fine-grained judges for multi-event TTA generation, explicitly assessing event presence and temporal order as indicators of instruction completeness and faithfulness. 
We further incorporate their feedback into preference-based training to improve the instruction-following capability of TTA models.

\section{Method}
% \subsection{ALLM-Judged Preference Optimization}
% ALLM-Judged Preference Optimization (LJPO)
\subsection{Overview: ALLM-Judged Preference Optimization}

We propose \textbf{A}LLM-\textbf{J}udged \textbf{P}reference \textbf{O}ptimization (AJPO), a training framework for improving instruction following in text-to-audio (TTA) generation. 
The key idea is to leverage audio-aware large language models (ALLMs) as judges to provide fine-grained, instruction-level evaluation of generated audio, and to use this evaluation to guide preference-based training of text-to-audio models.
Given a textual instruction specifying multiple sound events and their temporal relationships, a text-to-audio model first generates an audio clip. 
The ALLM then evaluates the generated audio with respect to the input instruction, explicitly assessing (i) whether each target sound event is present and (ii) whether the specified temporal order between events is satisfied. 
These judgments are performed at the instruction level, rather than relying on global audio–text similarity.
Based on the judgments produced by the ALLM, we construct preference pairs that distinguish instruction-following generations from those that violate event existence or temporal order constraints. 
These preference pairs are then used to optimize the text-to-audio model via direct preference optimization. 
By grounding preference construction in instruction-level judgments, AJPO directly targets multi-event generation and temporal order control, which are not explicitly addressed by conventional similarity-based objectives.
An overview of the proposed framework is illustrated in Figure~\ref{fig:overview}. 
Subsequent sections describe how we verify the reliability of ALLMs for instruction-level evaluation, detail the judging procedure, and present the preference construction and optimization process.

% \subsection{Verifying ALLMs as Judges for Instruction-Level Evaluation}
\subsection{Verifying ALLMs for Instruction-Level Evaluation}

% Since our framework relies on ALLMs as judges, a natural question is whether such models are capable of reliably evaluating instruction-level correctness in audio. 
% In this section, we examine the feasibility of using recent ALLMs to assess two key aspects required by our method: sound event existence and temporal ordering.
% To this end, we evaluate ALLMs on established audio understanding benchmarks that provide explicit annotations for sound events and temporal relationships. 
% For sound event existence, we consider datasets derived from AudioCaps~\cite{kim2019audiocaps}, where each audio clip is associated with a set of target sound events. 
% For temporal ordering, we use benchmarks such as CompA~\cite{ghoshcompa} and AudioTime~\cite{xie2025audiotime}, which are specifically designed to test a model’s ability to reason about the order of sound events in audio.
% Since existing benchmarks mainly involve two sound events with explicit ordering, we further construct controlled synthetic evaluation sets with more complex sound sequences. 
% Specifically, we concatenate audio clips from ESC-50~\cite{piczak2015dataset} to form evaluation sets with two, three, and four sound events, referred to as MultiEvent-Temporal-2, MultiEvent-Temporal-3, and MultiEvent-Temporal-4, respectively. 
% While these synthetic data represent idealized acoustic conditions, they allow us to systematically examine ALLM performance under controlled and increasingly challenging temporal configurations.

Since our framework relies on ALLMs as judges, we first validate whether they can reliably assess instruction-level correctness in audio.
Specifically, we evaluate their capabilities on two core aspects required by our method: sound event existence and temporal ordering.
We benchmark recent ALLMs on established audio understanding datasets with explicit annotations of sound events and temporal relations.
For sound event existence, we use datasets derived from AudioCaps~\cite{kim2019audiocaps}, where each clip is paired with annotated target events.
For temporal ordering, we adopt CompA~\cite{ghoshcompa} and AudioTime~\cite{xie2025audiotime}, which evaluate a model’s ability to reason about the order of sound events.
Because existing benchmarks mainly involve two events with explicit ordering, we further construct controlled synthetic evaluation sets with more complex sound sequences.
Specifically, we concatenate clips from ESC-50~\cite{piczak2015dataset} to create datasets containing two, three, and four events, denoted as MultiEvent-Temporal-2/3/4.
Although these synthetic settings represent idealized acoustic conditions, they enable a systematic validation of ALLM performance under increasingly challenging temporal structures.
We next describe the evaluation protocol used to assess ALLM performance on these tasks.
% For sound event existence, the ALLM is queried with a set of binary questions, each asking whether a specific target sound event is present in the audio. 
% Performance is evaluated using accuracy and exact match rate, which reflect both per-event correctness and holistic event coverage. 
% For temporal ordering, the ALLM is prompted to predict the relative order of sound events in the audio, and its predictions are compared with ground-truth sequences using exact match rate, pairwise accuracy, and Kendall’s tau.
% Our goal is not to optimize or fine-tune the ALLMs, but to assess whether off-the-shelf models already exhibit sufficient capability to serve as judges for instruction-level evaluation. 
% Taken together, these results suggest that recent ALLMs can reliably assess sound event existence and temporal ordering at the instruction level. 
% While not perfect, their performance is reasonably consistent to provide stable judgment signals, which form the basis of the instruction-level judging mechanism described next.
% For further verification of the ALLM-based judgments on TTA system outputs, we additionally conduct human evaluation to assess their correlation with human judgments.
For sound event existence, the ALLM is queried with a set of binary questions, each asking whether a specific target sound event is present in the audio. 
Performance is evaluated using accuracy and exact match rate, reflecting both per-event correctness and holistic event coverage. 
For temporal ordering, the ALLM predicts the relative order of sound events, and its outputs are compared with ground-truth sequences using exact match rate, pairwise accuracy, and Kendall’s tau.
Our goal is not to fine-tune the ALLMs, but to assess whether off-the-shelf models already provide sufficient capability to serve as judges for instruction-level evaluation. 
Taken together, the results show that recent ALLMs can reliably assess sound event existence and temporal ordering. 
Although not perfect, their performance is stable enough to provide consistent judgment signals, forming the basis of our instruction-level judging mechanism.
To further verify the reliability of ALLM-based judgments on TTA system outputs, we additionally conduct human evaluation to measure their correlation with human ratings.

\subsection{Instruction-Level Judging with ALLMs}

% Having verified that large audio-language models are capable of reliably assessing instruction-level correctness, we now describe how they are used as judges in our framework. 
We now describe how ALLMs are used as instruction-level judges in our framework. 
Given a generated audio clip and its corresponding textual instruction, the ALLM evaluates whether the audio satisfies the instruction with respect to two aspects: (i) sound event existence and (ii) temporal ordering.
% the ALLM evaluates whether the audio satisfies the instruction with respect to two aspects
% the ALLM evaluates whether the audio satisfies the instruction along two aspects
The model receives both the audio and instruction as input and produces judgments for these aspects. 
Rather than relying on implicit similarity scores, we prompt the model to output explicit decisions about event presence and temporal relationships.
% Given a generated audio clip and its corresponding textual instruction, the ALLM is tasked with explicitly evaluating whether the audio satisfies the instruction at a fine-grained level. 
% Specifically, the judging process focuses on two aspects: (i) sound event existence, which determines whether each target sound event specified in the instruction is present in the audio, and (ii) temporal ordering, which assesses whether the relative order of sound events matches the instruction.
% The ALLM receives both the audio and the textual instruction as input and produces structured judgments for these two aspects. 
% Rather than relying on implicit similarity scores, the model is prompted to provide explicit decisions regarding event presence and temporal relationships, enabling transparent and interpretable instruction-level evaluation.

\noindent\textbf{Sound Event Existence.}
Given a generated audio clip $a$ and instruction $t$, we determine whether each target sound event specified in $t$ is present in $a$. 
Let $\mathcal{E} = \{e_1, \ldots, e_N\}$ denote the set of target sound events. 
For each event $e_i \in \mathcal{E}$, the ALLM outputs a binary judgment $y_i \in \{0,1\}$ indicating its presence in the audio.
We aggregate these judgments into an instruction-level existence score:
\begin{equation}
s_{\text{exist}}(a, t)
= \frac{1}{|\mathcal{E}|} \sum_{i=1}^{|\mathcal{E}|} y_i ,
\end{equation}
which measures the proportion of required events present in the audio. 
A higher score indicates better adherence to the sound event specification. 
Although a strict exact-match criterion can also be derived, we primarily use the averaged score for preference construction due to its finer granularity.

\noindent\textbf{Temporal Ordering.}
Given $a$ and $t$, the ALLM predicts the relative order of the target sound events. 
The model outputs a ranking $\hat{\pi}$ over $\mathcal{E}$, where $\hat{\pi}(e_i)$ denotes the predicted occurrence rank of event $e_i$. 
This formulation supports both total and partial order constraints.
Let $\pi$ denote the target ordering implied by the instruction. 
We compute several temporal metrics for analysis, including exact match rate, pairwise accuracy, and Kendall’s tau. 
For preference construction, we adopt Kendall’s tau as the temporal order score:
\begin{equation}
s_{\text{order}}(a, t)
= \tau(\hat{\pi}, \pi),
\end{equation}
where $\tau(\cdot, \cdot) \in [-1, 1]$ denotes rank correlation. 
A higher score indicates better alignment with the specified temporal order.
The resulting existence and temporal scores form structured instruction-level judgments, which are used to construct preference data and optimize TTA models.

\subsection{Preference Construction and Optimization}

We construct preference data directly from the instruction-level judgments produced by ALLMs. 
The sound event existence score and temporal order score provide complementary signals of instruction adherence, enabling structured pairwise supervision.
For each instruction $t$, we sample candidate audio generations $\{a^{(k)}\}_{k=1}^{K}$ from the current TTA model and compute the corresponding judgment scores.
Let $s_{\text{exist}}(a,t) \in [0,1]$ denote the sound event existence score and $s_{\text{order}}(a,t) \in [-1,1]$ denote the temporal order score (Kendall’s $\tau$).
We construct preference pairs $(a^{+}, a^{-})$ by selecting a high-quality anchor sample and contrasting it with rejected samples that violate instruction-level constraints.
An anchor sample $a^{+}$ must satisfy
\begin{equation}
s_{\text{exist}}(a^{+}, t) = 1.0
\quad \text{and} \quad
s_{\text{order}}(a^{+}, t) = 1.0,
\end{equation}
ensuring perfect event coverage and temporal correctness.
We consider two types of rejected samples $a^{-}$. 
The first contains all target events but violates temporal order:
\begin{equation}
s_{\text{exist}}(a^{-}, t) = 1.0
\quad \text{and} \quad
s_{\text{order}}(a^{-}, t) < 1.0.
\end{equation}
The second exhibits incomplete event generation:
\begin{equation}
s_{\text{exist}}(a^{-}, t) < 1.0.
\end{equation}
These structured preferences distinguish temporal-order violations from missing-event failures and provide diagnostic supervision for direct preference optimization.

\subsection{Sound Scene Story Benchmark}

Existing TTA benchmarks mainly involve short and simple instructions, often containing only one or two sound events. 
For example, an AudioCaps caption may read:
\textit{``Rain is falling continuously.''}
Such captions describe a single event without temporal transitions or compositional structure.
In contrast, an instruction from our proposed benchmark may resemble a short auditory narrative, such as:
\textit{``Rain begins to fall steadily, a car drives past splashing through puddles, and moments later distant thunder rumbles as the rain continues.''}
Unlike the former, this instruction requires reasoning over multiple interacting sound events and their temporal progression. 
These differences suggest that existing benchmarks might not sufficiently evaluate complex instruction following involving multi-event and temporal reasoning.
% To address this limitation, we introduce the Sound Scene Story Benchmark (S3Bench), designed to evaluate instruction following under more expressive and compositional scenarios. 
To address this limitation, we design the Sound Scene Story Benchmark (S3Bench) to evaluate instruction following under more expressive and compositional scenarios. 
S3Bench consists of narrative-style instructions describing sound scenes with multiple events and explicit temporal structure, rather than fixed templates.
The benchmark covers diverse temporal patterns, including sequential progressions (e.g., $e_1 \rightarrow e_2 \rightarrow e_3$) and partial-order cases where events overlap before transitioning (e.g., $e_1$ and $e_2$ occurring together before $e_3$). 
This design evaluates both strict ordering and overlapping temporal reasoning.
S3Bench contains 1,200 instances: 300 with two events, 500 with three events, and 200 with four events. 
Additionally, 200 instances explicitly involve overlapping events. 
These configurations allow evaluation under varying levels of compositional complexity.
To construct S3Bench, we first sample two to four sound event labels from a predefined vocabulary derived from ESC-50 and AudioCaps. 
We then prompt a large language model to generate narrative instructions conditioned on these events, using manually designed in-context examples to guide coherent and temporally explicit descriptions. 
This procedure enables scalable construction with controlled linguistic consistency.
S3Bench is used solely for evaluation and is not included in model training.

\begin{table}[tbp]
\centering
\scriptsize
\setlength{\tabcolsep}{4pt}
\caption{
Performance of ALLMs on sound event existence and temporal order evaluation.
Accuracy metrics are in \%.
}
\begin{tabular}{l cc ccc c}
\toprule
& \multicolumn{2}{c}{\textbf{Existence}} 
& \multicolumn{3}{c}{\textbf{Temporal Order}} 
& \textbf{OVL} \\
\cmidrule(lr){2-3} \cmidrule(lr){4-6} \cmidrule(lr){7-7}
\textbf{Model} 
& \textbf{EM } & \textbf{Micro } 
& \textbf{EM } & \textbf{Pair. } & $\boldsymbol{\tau}$ 
& \textbf{Joint } \\
\midrule

\multicolumn{7}{l}{\textit{AudioCaps-test}} \\
Qwen2.5-Omni-7B 
& \textbf{81.7} & \textbf{91.9} & 75.4 & 83.3 & 0.67 & \textbf{51.8} \\
Qwen2.5-Omni-3B 
& \underline{73.6} & \underline{87.5} & 76.7 & 83.7 & 0.68 & \underline{45.7} \\
Audio Flamingo 3 
& 42.7 & 66.0 & 80.0 & 83.9 & 0.68 & 25.9 \\
% CLAP
% & -- & -- & 46.7 & 57.2 & 0.12 & -- \\
\midrule

\multicolumn{7}{l}{\textit{CompA}} \\
Qwen2.5-Omni-7B 
& \textbf{72.4} & \textbf{85.1} & 89.3 & 89.3 & 0.79 & \textbf{64.7} \\
Qwen2.5-Omni-3B 
& \underline{65.8} & \underline{80.8} & 88.1 & 88.1 & 0.76 & \underline{58.0} \\
Audio Flamingo 3 
& 37.5 & 61.3 & 89.1 & 89.1 & 0.78 & 33.2 \\
% CLAP
% & -- & -- & 48.0 & 48.0 & -0.03 & -- \\
\midrule

\multicolumn{7}{l}{\textit{AudioTime}} \\
Qwen2.5-Omni-7B 
& \textbf{95.0} & \textbf{97.5} & 99.4 & 99.4 & 0.99 & \textbf{94.4} \\
Qwen2.5-Omni-3B 
& \underline{88.0} & \underline{94.0} & 98.4 & 98.4 & 0.97 & \underline{86.6} \\
Audio Flamingo 3 
& 70.6 & 84.5 & 95.5 & 95.5 & 0.91 & 67.4 \\
% CLAP
% & -- & -- & 49.0 & 49.0 & -0.01 & -- \\
\midrule

\multicolumn{7}{l}{\textit{MultiEvent-Temporal-2}} \\
Qwen2.5-Omni-7B 
& \textbf{89.5} & \textbf{94.7} & 99.7 & 99.7 & 0.99 & \textbf{89.1}
\\
Qwen2.5-Omni-3B 
& \underline{86.5} & \underline{93.0} & 99.0 & 99.0 & 0.98 & \underline{85.3}
\\
Audio Flamingo 3
& 53.8 & 73.6 & 97.6 & 97.6 & 0.95 & 52.3
\\
% CLAP
% & -- & -- & 49.6 & 49.6 & -0.01 & -- \\
\midrule

\multicolumn{7}{l}{\textit{MultiEvent-Temporal-3}} \\
Qwen2.5-Omni-7B 
& \textbf{83.2} & \textbf{94.0} & 93.2 & 97.7 & 0.95 & \textbf{77.5} \\
Qwen2.5-Omni-3B 
& \underline{78.2} & \underline{92.1} & 84.7 & 94.7 & 0.89 & \underline{66.2} \\
Audio Flamingo 3 
& 38.4 & 72.9 & 86.2 & 93.5 & 0.87 & 33.1 \\
% CLAP
% & -- & -- & 15.8 & 49.4 & -0.01 & -- \\
\midrule

\multicolumn{7}{l}{\textit{MultiEvent-Temporal-4}} \\
Qwen2.5-Omni-7B 
& \textbf{70.9} & \textbf{92.0} & 81.2 & 96.6 & 0.93 & \textbf{57.6} \\
Qwen2.5-Omni-3B 
& \underline{69.3} & \underline{91.6} & 62.3 & 91.9 & 0.84 & \underline{43.2} \\
Audio Flamingo 3 
& 25.7 & 71.5 & 61.5 & 88.1 & 0.76 & 15.8 \\
% CLAP
% & -- & -- & 4.7 & 50.7 & 0.01 & -- \\
\bottomrule
\end{tabular}
\label{tab:lalm-instruction-eval}
\end{table}
% \vspace{-15pt}
\begin{table}[tbp]
\centering
\small
\setlength{\tabcolsep}{6pt}
% \caption{
% Human verification results of ALLM-based instruction-level judgments on TTA-generated audio samples. 
% For sound event existence, agreement is reported at the event level, while correlation is computed over instance-level existence scores. 
% For temporal order, Kendall’s $\tau$ is computed between human and ALLM-predicted orderings, evaluated only when all target events are judged present.
% }
\caption{
Human verification of ALLM-based instruction-level judgments on TTA outputs.
For existence, we report event-level agreement and instance-level score correlation.
For temporal order, Kendall's $\tau$ is computed between human and ALLM orderings over samples where all target events are judged present.
}
\begin{tabular}{lcc}
\toprule
\textbf{Metric} & \textbf{Existence} & \textbf{Temporal Order} \\
\midrule
Agreement (\%) 
& 89.8 
& 93.5 \\
\midrule
Pearson $r$ 
& 0.85 
& - \\
Spearman $\rho$ 
& 0.83 
& - \\
Kendall's $\tau$ 
& 0.76 
& 0.93 \\
\bottomrule
\end{tabular}
\label{tab:human-verification}
\end{table}

% \caption{
% % Human verification results of ALLM-based instruction-level judgments on TTA-generated audio samples.
% % For sound event existence, agreement is measured at the event level, while correlation is computed based on instance-level existence scores.
% % For temporal order, agreement is measured using Kendall’s $\tau$ between human and ALLM orderings, evaluated only when all target events are judged present by humans.
% Human verification results of ALLM-based instruction-level judgments on TTA-generated audio samples. 
% For sound event existence, agreement is reported at the event level, while correlation is computed over instance-level existence scores. 
% For temporal order, Kendall’s $\tau$ is computed between human and ALLM-predicted orderings, evaluated only when all target events are judged present.
% }

\section{Experimental Setup}

\subsection{Evaluation Benchmarks}

We use different benchmarks to support two evaluation goals: 
(1) verifying whether ALLMs are sufficiently capable to serve as instruction-level judges, and 
(2) evaluating the instruction-following performance of TTA systems under varying levels of complexity.

\noindent\textbf{Verifying ALLMs as Judges.}
To assess ALLMs, we evaluate them on benchmarks with explicit annotations of sound event existence and temporal ordering. 
AudioCaps-test is used to evaluate sound event existence. 
For temporal reasoning, we use CompA and AudioTime, which focus on identifying relative event order in audio. 
In addition, we construct controlled synthetic multi-event evaluation sets by concatenating ESC-50 audio clips, forming MultiEvent-Temporal-2/3/4, which contain two, three, and four sound events, respectively. 
These synthetic sets allow controlled analysis of temporal reasoning behavior.

\noindent\textbf{Evaluating Text-to-Audio Generation.}
To evaluate TTA systems, we use benchmarks designed to assess instruction following in generated audio. 
AudioCaps-test and the text instructions from the MultiEvent-Temporal series are used as prompts for generation. 
Note that the corresponding synthetic audio clips are used only for evaluating ALLMs, not for TTA evaluation. 
Finally, we evaluate generalization using the Sound Scene Story Benchmark (S3Bench), which contains narrative-style multi-event instructions with explicit temporal structure.

\subsection{Evaluation Metrics}

We use a unified set of metrics to evaluate both ALLMs and TTA systems. 
While the metrics are identical in form, they are applied to different targets: ALLMs are evaluated based on the accuracy of their judgments, whereas TTA systems are evaluated on whether their generated audio satisfies the instruction.
% evaluated based on the accuracy of their judgments
\noindent\textbf{Evaluation of ALLMs.}
To evaluate ALLMs as judges, we apply sound event existence and temporal order metrics to their predictions on audio understanding benchmarks.
For sound event existence, we report \textit{Exact Match} and \textit{Micro Accuracy}, measuring strict instruction satisfaction and per-event correctness.
For temporal ordering, we report \textit{Exact Match}, \textit{Pairwise Accuracy}, and \textit{Kendall's Tau}, capturing sequence-level correctness and rank consistency.

\noindent\textbf{Evaluation of TTA Models.}
For TTA systems, the same metrics are applied to generated audio to measure instruction-following performance.
These metrics quantify whether the specified sound events are present and whether their temporal relationships are satisfied.
In addition, we report \textit{Joint Accuracy}, which requires both sound event existence and temporal ordering to be correct.
% We also report standard audio generation metrics, including FAD, FD, KL, IS, and CLAPScore, to assess audio quality and overall audio-text alignment.
We also report standard audio generation metrics, including FAD, FD, KL, IS, and CLAPScore.
In the following tables, for brevity, \textit{EM} denotes Exact Match, \textit{Micro} denotes micro accuracy, \textit{$\tau$} denotes Kendall’s tau, and \textit{Joint} denotes Joint Accuracy.

\noindent\textbf{Evaluation Condition.}
% Temporal order metrics are computed only when sound event existence Exact Match equals 1.0.
Temporal-order metrics are computed only for samples whose sound-event-existence exact-match score is 1.0.
Thus, temporal performance should be interpreted as a conditional measure, reflecting ordering accuracy given complete event presence.
This ensures that the temporal metric is meaningful, as ordering is only well-defined when all relevant events are correctly identified.

\subsection{Training Data Construction}

We construct preference training data using captions from the AudioCaps training set and automatically generated instruction templates. 
From AudioCaps, we select captions that involve multiple sound events with implicit temporal relationships, providing diverse multi-event instructions in natural language.
We also construct temporally explicit instruction templates using sound event labels from ESC-50. 
An example template is of the form: 
``It starts with the sound of $e_1$, shifts to $e_2$, and ends with $e_3$.'' 
These templates explicitly encode temporal order and are used only for training. 
To prevent template overlap with evaluation benchmarks, the training templates differ from those used in the MultiEvent-Temporal-2/3/4 benchmarks.
In total, we construct 60k training samples for preference-based optimization. 
Each sample consists of a textual instruction paired with multiple candidate audio generations, which are evaluated by ALLMs to form preference pairs.

\subsection{Baselines and Training Details}

We consider baselines from both audio understanding and text-to-audio generation perspectives. 
For ALLMs used as judges, we evaluate Qwen2.5-Omni~\cite{xu2025qwen2} (3B and 7B) and Audio Flamingo 3~\cite{goel2025audio}. 
For TTA generation, we include AudioLDM2-full-large~\cite{audioldm2-2024taslp}, EzAudio-XL~\cite{hai2024ezaudio}, Stable-Audio-Open~\cite{evans2025stable}, Tango2~\cite{majumder2024tango}, and TangoFlux-base and TangoFlux-CRPO~\cite{hung2024tangofluxsuperfastfaithful}. 
These models cover both diffusion- and flow-based architectures. 
We follow the recommended configurations in the original papers, including classifier-free guidance scale and denoising steps.
For preference-based training, TangoFlux-base serves as the base model for both candidate generation and direct preference optimization. 
We compare two training objectives: direct preference optimization (DPO) and supervised fine-tuning (SFT). 
For SFT, the model is trained using only preferred samples, discarding rejected ones. 
% All other training configurations are identical across objectives for fair comparison.
All models are trained on 8 NVIDIA A100 GPUs. 
We use a learning rate of $1\times10^{-4}$ with a global batch size of 128 and apply a linear learning rate scheduler with 500 warm-up steps.
Unless otherwise specified (e.g., in the DPO iteration study), all experiments are conducted with a single training iteration.

\subsection{Human Verification Setup}

To further assess the reliability of ALLMs as instruction-level judges in practical TTA settings, we conduct a human verification study. 
Three human annotators evaluate 150 randomly sampled audio clips generated by multiple representative TTA systems, covering instructions with multiple sound events and explicit temporal constraints.
Annotators judge the presence of each target sound event and provide a full temporal ordering only when all events are confirmed present. 
For sound event existence, final human labels are determined by majority voting, and we report event-level agreement and instance-level correlation with ALLM judgments. 
For temporal ordering, we compare full event sequences using Kendall’s tau, computed only when all target events are confirmed present.

\begin{table*}[tbp]
\centering
\small
% \scriptsize
\setlength{\tabcolsep}{5pt}
\caption{
Evaluation on AudioCaps-test.
% All accuracy-based metrics are reported in percentage.
Accuracy-based metrics are reported as percentages.
% Temporal order is evaluated on a subset of AudioCaps-test where all target sound events are present.
}
\begin{tabular}{l cc ccc c ccccc}
\toprule
& \multicolumn{2}{c}{\textbf{Event Existence}} 
& \multicolumn{3}{c}{\textbf{Temporal Order}} 
& \textbf{Overall} 
& \multicolumn{5}{c}{\textbf{Common Metrics}} \\
\cmidrule(lr){2-3}
\cmidrule(lr){4-6}
\cmidrule(lr){7-7}
\cmidrule(lr){8-12}
\textbf{Model}
& \textbf{EM} & \textbf{Micro}
& \textbf{EM} & \textbf{Pairwise} & $\boldsymbol{\tau}$
& \textbf{Joint}
& \textbf{FAD} & \textbf{FD} & \textbf{KL} & \textbf{IS} & \textbf{CLAP} \\
\midrule

AudioLDM2
& 52.8 & 74.7
& 56.8 & 63.0 & 0.26
& 20.6
& \textbf{2.08} & 34.28 & 1.69 & 7.34 & 0.264 \\

EzAudio
& 84.3 & 93.3
& 72.5 & 81.4 & 0.63
& 51.8
& 3.27 & \textbf{15.40} & \underline{1.23} & 11.02 & \textbf{0.414} \\

Stable-Audio
& 42.8 & 69.8
& 55.9 & 61.8 & 0.24
& 15.3
& 4.83 & 41.56 & 2.25 & 9.62 & 0.255 \\

Tango2
& 83.9 & 93.3
& 77.4 & 83.9 & 0.68
& 56.3
& 2.71 & 21.01 & \textbf{1.16} & 8.76 & 0.373 \\

TangoFlux-base
& \underline{84.8} & \underline{93.5}
& 85.2 & 92.6 & 0.89
& 67.1
& 3.37 & \underline{19.92} & \textbf{1.16} & 11.42 & 0.365 \\

TangoFlux-CRPO
& \textbf{87.4} & \textbf{94.7}
& 85.7 & 90.7 & 0.81
& \underline{67.4}
& \underline{2.34} & 20.23 & \textbf{1.16} & \textbf{11.96} & \underline{0.397} \\

\rowcolor{lightgreen} \textbf{Ours}
& \textbf{87.4} & \textbf{94.7}
& 89.1 & 92.8 & 0.86
& \textbf{71.0}
& 3.31 & 20.07 & \textbf{1.16} & \underline{11.54} & 0.375 \\

\bottomrule
\end{tabular}
\label{tab:audiocaps-instruction-eval}
\end{table*}

\begin{table}[t]
\centering
\scriptsize
\setlength{\tabcolsep}{5pt}
% \vspace{+15pt}
% \begin{tabular}{l cc ccc c cc}
\caption{
Instruction-level evaluation on multi-event temporal benchmarks.
Accuracy metrics are reported in \%.
}
\begin{tabular}{l cc ccc c}
\toprule
& \multicolumn{2}{c}{\textbf{Event Existence}}
& \multicolumn{3}{c}{\textbf{Temporal Order}}
& \textbf{OVL}
% & \multicolumn{2}{c}{\textbf{Human}}
\\
\cmidrule(lr){2-3}
\cmidrule(lr){4-6}
\cmidrule(lr){7-7}
% \cmidrule(lr){8-9}
\textbf{Model}
& \textbf{EM} & \textbf{Micro}
& \textbf{EM} & \textbf{Pair.} & $\boldsymbol{\tau}$
& \textbf{Joint} 
% & \textbf{REL} & \textbf{OVL}
\\
\midrule
% MultiEvent-Temporal-2
\multicolumn{7}{l}{\textit{MultiEvent-Temporal-2}} \\
% AudioLDM2-full-Large 
AudioLDM2
& 21.0 & 54.4 & 55.7 & 55.7 & 0.11 & 11.7 \\
% EzAudio-XL 
EzAudio
& 63.8 & 81.3 & 72.0 & 72.0 & 0.44 & 46.0 \\
% Stable-Audio-Open    
Stable-Audio
& 27.0 & 62.0 & 57.6 & 57.6 & 0.15 & 15.5 \\
Tango2                
& 67.5 & 83.1 & 68.6 & 68.6 & 0.37 & 46.0 \\
TangoFlux-base       
& 77.9 & 88.9 & 92.4 & 92.4 & 0.85 & 72.0 \\
TangoFlux-CRPO    
& \underline{85.9} & \underline{92.9} & 94.3 & 94.3 & 0.89 & \underline{81.0} \\
\rowcolor{lightgreen} \textbf{Ours}         
& \textbf{89.0} & \textbf{94.5} & 96.1 & 96.1 & 0.92 & \textbf{85.5} \\
\midrule

\multicolumn{7}{l}{\textit{MultiEvent-Temporal-3}} \\
AudioLDM2
& 3.5  & 40.0 & 5.7  & 49.5 & -0.01 & 0.2 \\
EzAudio
& 35.2 & 72.0 & 30.7 & 66.5 & 0.33 & 10.8 \\
Stable-Audio
& 8.2  & 49.8 & 20.7 & 53.7 & 0.07 & 1.7 \\
Tango2                
& 40.2 & 75.2 & 27.3 & 63.4 & 0.27 & 11.0 \\
TangoFlux-base      
& 48.3 & 80.4 & 67.9 & 88.4 & 0.77 & 32.8 \\
TangoFlux-CRPO  
& \underline{61.4} & \underline{85.8} & 65.2 & 88.0 & 0.76 & \underline{40.0} \\
\rowcolor{lightgreen} \textbf{Ours}         
& \textbf{69.1} & \textbf{88.7} & 79.6 & 92.9 & 0.86 & \textbf{55.0} \\
\midrule

\multicolumn{7}{l}{\textit{MultiEvent-Temporal-4}} \\
AudioLDM2
& 2.3  & 37.0 & 0.0  & 39.9 & -0.20 & 0.0 \\
EzAudio   
& 19.0 & 67.3 & 9.5  & 63.7 & 0.27 & 1.8 \\
Stable-Audio
& 3.0  & 44.1 & 6.7  & 52.8 & 0.06 & 0.2 \\
Tango2     
& 22.2 & 69.6 & 10.8 & 62.4 & 0.25 & 2.4 \\
TangoFlux-base 
& 24.0 & 72.8 & 31.7 & 81.9 & 0.64 & 7.6 \\
TangoFlux-CRPO
& \underline{33.2} & \underline{78.6} & 34.9 & 83.8 & 0.68 & \underline{11.6} \\
\rowcolor{lightgreen} \textbf{Ours}         
& \textbf{41.3} & \textbf{81.4} & 44.6 & 87.3 & 0.75 & \textbf{18.4} \\
\midrule

\multicolumn{7}{l}{\textit{Sound Scene Story Bench}} \\
AudioLDM2
& 4.4  & 36.3 & 41.5 & 50.0 & 0.00 & 1.7 \\
EzAudio
& 27.8 & 65.5 & 54.2 & 68.0 & 0.36 & 14.5 \\
Stable-Audio
& 11.8 & 48.8 & 41.6 & 54.5 & 0.09 & 4.6 \\
Tango2
& 36.1 & 70.9 & 49.3 & 66.7 & 0.34 & 17.0 \\
TangoFlux-base
& 43.7 & 75.3 & 82.3 & 91.9 & 0.84 & 35.3 \\
TangoFlux-CRPO
& \underline{57.0} & \underline{82.2} & 80.6 & 92.3 & 0.85 & \underline{45.4} \\
\rowcolor{lightgreen} \textbf{Ours}
& \textbf{59.2} & \textbf{83.2} & 85.1 & 94.4 & 0.89 & \textbf{49.9} 
\\
\bottomrule
\end{tabular}
\label{tab:multi-event-eval}
\end{table}

\begin{table*}[tbp]
\centering
\small
\setlength{\tabcolsep}{5pt}
\caption{
Ablation study on AudioCaps-test.
% All accuracy-based metrics are reported in percentage.
% Kendall’s $\tau$ is reported as a correlation coefficient.
% Macro accuracy is omitted for clarity.
% Temporal order is evaluated on a subset of AudioCaps-test where all target sound events are present.
Gray indicates the base model. 
Orange indicates other existing preference training data. 
% Green indicates our source preference data constructed using different pairing strategies or optimization objectives.
Green indicates variants using our preference data with different pairing strategies or optimization objectives.
% Accuracy-based metrics are reported as percentages.
% Accuracy metrics are reported in \%.
}
\begin{tabular}{l cc ccc c ccccc cc}
\toprule
& \multicolumn{2}{c}{\textbf{Event Existence}} 
& \multicolumn{3}{c}{\textbf{Temporal Order}} 
& \textbf{OVL} 
& \multicolumn{5}{c}{\textbf{Common Metrics}} 
& \multicolumn{2}{c}{\textbf{Subjective}} 
\\
\cmidrule(lr){2-3}
\cmidrule(lr){4-6}
\cmidrule(lr){7-7}
\cmidrule(lr){8-12}
\cmidrule(lr){13-14}
\textbf{Method}
& \textbf{EM} & \textbf{Micro}
& \textbf{EM} & \textbf{Pairwise} & $\boldsymbol{\tau}$
& \textbf{Joint}
& \textbf{FAD} & \textbf{FD} & \textbf{KL} & \textbf{IS} & \textbf{CLAP} 
& \textbf{REL} & \textbf{OVL}
\\
\midrule

\rowcolor{lightgray} TangoFlux Base
& 84.8 & 93.5
& 85.2 & 92.6 & 0.89
& 67.1
& 3.37 & \textbf{19.92} & \textbf{1.16} & 11.42 & 0.365 
& 3.39 & \underline{3.38}
\\

\rowcolor{lightorange} Audio Alpaca
& \underline{86.9} & 94.6
& 88.6 & 93.0 & 0.86
& 68.8
& 3.47 & 21.53 & \underline{1.17} & \textbf{11.77} & \underline{0.377} 
& -- & --
\\

\rowcolor{lightorange} Baton
& 85.2 & 94.5
& 89.5 & 93.4 & 0.87
& \underline{69.1}
& 3.38 & 20.17 & \textbf{1.16} & 11.36 & 0.367 
& -- & --
\\

\rowcolor{lightgreen} CLAP-DPO
& 86.0 & 93.8
& 87.9 & 91.8 & 0.84
& 68.3
& 3.35 & 20.52 & \underline{1.17} & \underline{11.61} & \textbf{0.379} 
& \underline{3.55} & 3.32
\\

\rowcolor{lightgreen} ALLM-CAP
& 85.7 & \textbf{94.8}
& 89.5 & 92.5 & 0.85
& 68.0
& \textbf{3.28} & 20.48 & \underline{1.17} & 11.45 & 0.373
& -- & --
\\

\rowcolor{lightgreen} ALLM-SFT
& 85.3 & 93.4
& 88.6 & 93.1 & 0.86
& 67.1
& 32.43 & 26.45 & 1.31 & 10.51 & 0.359 
& -- & --
\\

% \midrule

\rowcolor{lightgreen} ALLM-DPO
& \textbf{87.4} & \underline{94.7}
& 89.1 & 92.8 & 0.86
& \textbf{71.0}
& \underline{3.31} & \underline{20.07} & \textbf{1.16} & 11.54 & 0.375 
& \textbf{4.28} & \textbf{3.39}
\\

% LALM-DPO (2)
% & 90.8 & 95.8
% & 88.8 & 93.7 & 0.87
% & 74.7
% & 3.26 & 20.18 & 1.17 & 11.26 & 0.377
% & -- & --
% \\

% LALM-DPO (3)
% & 90.5 & 96.6
% & 90.8 & 94.3 & 0.89
% & 76.0
% & 3.31 & 20.79 & 1.19 & 10.95 & 0.377
% & -- & --
% \\

% LALM-DPO (4)
% & 90.3 & 96.3
% & 91.0 & 94.5 & 0.89
% & 76.3
% & 3.50 & 20.61 & 1.22 & 10.95 & 0.374
% & -- & --
% \\

% LALM-DPO (5)
% & 91.4 & 96.8
% & 90.7 & 94.2 & 0.88
% & 76.6
% & 3.78 & 20.58 & 1.23 & 10.70 & 0.371
% & -- & --
% \\

% LALM-DPO+CRPO
% & 88.4 & 95.8
% & 87.7 & 91.7 & 0.83
% & 70.5
% & 3.17 & 19.35 & 1.17 & 11.55 & 0.379
% & -- & --
% \\

\bottomrule
\end{tabular}
\label{tab:audiocaps-ablation}
\end{table*}
\begin{table}[t]
\centering
\scriptsize
\setlength{\tabcolsep}{5pt}
% \vspace{+10pt}
\caption{
Ablation Study on multi-event temporal benchmarks.
% All accuracy-based metrics are reported in percentage.
Accuracy-based metrics are reported as percentages.
}
\begin{tabular}{l cc ccc c}
\toprule
& \multicolumn{2}{c}{\textbf{Event Existence}}
& \multicolumn{3}{c}{\textbf{Temporal Order}}
& \textbf{OVL} \\
\cmidrule(lr){2-3}
\cmidrule(lr){4-6}
\cmidrule(lr){7-7}
\textbf{Method}
& \textbf{EM} & \textbf{Micro}
& \textbf{EM} & \textbf{Pair.} & $\boldsymbol{\tau}$
& \textbf{Joint} \\
\midrule

\multicolumn{7}{l}{\textit{MultiEvent-Temporal-2}} \\
\rowcolor{lightgray} TangoFlux-base       
& 77.9 & 88.9 & 92.4 & 92.4 & 0.85 & 72.0 \\
\rowcolor{lightorange} Audio Alpaca 
& 82.1 & \underline{91.0} & 95.6 & 95.6 & 0.91 & 78.5 \\
\rowcolor{lightorange} Baton 
& 77.4 & 88.5 & 93.9 & 93.9 & 0.88 & 72.7 \\
\rowcolor{lightgreen} CLAP-DPO 
& 82.1 & \underline{91.0} & 93.4 & 93.4 & 0.87 & 76.7 \\
\rowcolor{lightgreen} ALLM-CAP
& 78.7 & 89.2 & 94.8 & 94.8 & 0.90 & 74.6 
\\
\rowcolor{lightgreen} ALLM-SFT 
& \underline{82.2} & 90.8 & 96.0 & 96.0 & 0.92 & \underline{78.9} \\
\rowcolor{lightgreen} ALLM-DPO 
& \textbf{89.0} & \textbf{94.5} & 96.1 & 96.1 & 0.92 & \textbf{85.5} 
\\
% ALLM-DPO (2)        & 93.0 & 96.5 & 98.3 & 98.3 & 0.97 & 91.4 \\
% ALLM-DPO (3)        & 93.8 & 96.8 & 98.8 & 98.8 & 0.98 & 92.7 \\
% ALLM-DPO (4)        & 94.3 & 97.1 & 98.8 & 98.8 & 0.98 & 93.2 \\
% ALLM-DPO (5)        & 94.2 & 97.0 & 98.8 & 98.8 & 0.98 & 93.1 \\
% DPO+CRPO & 89.5 & 94.8 & 96.3 & 96.3 & 0.93 & 86.2 \\

\midrule

\multicolumn{7}{l}{\textit{MultiEvent-Temporal-3}} \\
\rowcolor{lightgray} TangoFlux-base      
& 48.3 & 80.4 & 67.9 & 88.4 & 0.77 & 32.8 \\
\rowcolor{lightorange} Audio Alpaca 
& \underline{57.2} & \underline{83.9} & 71.5 & 90.0 & 0.80 & \underline{40.9} \\
\rowcolor{lightorange} Baton 
& 49.8 & 80.9 & 69.3 & 89.3 & 0.79 & 34.5 \\
\rowcolor{lightgreen} CLAP-DPO 
& 54.4 & 83.1 & 68.6 & 88.9 & 0.78 & 37.3 \\
\rowcolor{lightgreen} ALLM-CAP
& 53.6 & 82.6 & 70.7 & 89.7 & 0.79 & 37.9 \\
\rowcolor{lightgreen} ALLM-SFT 
& 51.5 & 81.1 & 72.0 & 90.4 & 0.81 & 37.1 \\
\rowcolor{lightgreen} ALLM-DPO 
& \textbf{69.1} & \textbf{88.7} & 79.6 & 92.9 & 0.86 & \textbf{55.0} \\
% ALLM-DPO (2)        & 80.0 & 92.8 & 86.9 & 95.5 & 0.91 & 69.5 \\
% ALLM-DPO (3)        & 81.3 & 93.4 & 90.5 & 96.6 & 0.93 & 73.6 \\
% ALLM-DPO (4)        & 81.9 & 93.7 & 92.6 & 97.2 & 0.95 & 75.8 \\
% ALLM-DPO (5)        & 82.0 & 93.5 & 93.3 & 97.4 & 0.95 & 76.5 \\
% DPO+CRPO & 66.9 & 87.9 & 79.7 & 92.9 & 0.86 & 53.3 \\
\midrule

\multicolumn{7}{l}{\textit{MultiEvent-Temporal-4}} \\
\rowcolor{lightgray} TangoFlux-base 
& 24.0 & 72.8 & 31.7 & 81.9 & 0.64 & 7.6 \\
\rowcolor{lightorange} Audio Alpaca 
& 29.3 & 75.6 & 39.9 & 85.6 & 0.71 & \underline{11.7} \\
\rowcolor{lightorange} Baton        
& 24.2 & 73.1 & 34.3 & 82.6 & 0.65 & 8.3 \\
\rowcolor{lightgreen} CLAP-DPO 
& \underline{31.0} & \underline{76.4} & 32.9 & 84.0 & 0.68 & 10.2 \\
\rowcolor{lightgreen} ALLM-CAP 
& 26.5 & 74.6 & 35.9 & 83.6 & 0.67 & 9.5 
\\
\rowcolor{lightgreen} ALLM-SFT 
& 24.4 & 72.7 & 41.4 & 86.3 & 0.73 & 10.1 \\
\rowcolor{lightgreen} ALLM-DPO 
& \textbf{41.3} & \textbf{81.4} & 44.6 & 87.3 & 0.75 & \textbf{18.4} \\
% ALLM-DPO (2)        & 55.1 & 86.7 & 54.3 & 90.5 & 0.81 & 29.9 \\
% ALLM-DPO (3)        & 58.9 & 88.1 & 63.8 & 92.6 & 0.85 & 37.6 \\
% ALLM-DPO (4)        & 64.2 & 93.7 & 66.2 & 93.1 & 0.86 & 42.5 \\
% ALLM-DPO (5)        & 63.7 & 89.4 & 69.5 & 93.9 & 0.88 & 44.3 \\
% DPO+CRPO & 36.8 & 79.7 & 46.5 & 87.4 & 0.75 & 17.1 \\

\midrule

\multicolumn{7}{l}{\textit{Sound Scene Story Bench}} \\
\rowcolor{lightgray} TangoFlux-base
& 43.7 & 75.3 & 82.3 & 91.9 & 0.84 & 35.3 \\
\rowcolor{lightorange} Audio Alpaca 
& \underline{53.2} & \underline{79.9} & 83.9 & 93.7 & 0.87 & \underline{44.1} \\
\rowcolor{lightorange} Baton 
& 43.6 & 75.8 & 82.9 & 92.1 & 0.84 & 35.5 \\
\rowcolor{lightgreen} CLAP-DPO 
& 51.7 & 79.6 & 81.2 & 92.0 & 0.84 & 41.3 \\
\rowcolor{lightgreen} ALLM-CAP
& 49.4 & 79.6 & 83.6 & 93.3 & 0.87 & 40.8 \\
\rowcolor{lightgreen} ALLM-SFT 
& 50.9 & 78.9 & 85.8 & 94.8 & 0.90 & 43.2 \\
\rowcolor{lightgreen} ALLM-DPO 
& \textbf{59.2} & \textbf{83.2} & 85.1 & 94.4 & 0.89 & \textbf{49.9} \\
% ALLM-DPO (2)        & 68.8 & 88.6 & 89.9 & 96.6 & 0.93 & 61.5 \\
% ALLM-DPO (3)        & 74.3 & 90.7 & 92.0 & 97.5 & 0.95 & 68.0 \\
% ALLM-DPO (4)        & 77.0 & 91.9 & 93.4 & 97.9 & 0.96 & 71.7 \\
% ALLM-DPO (5)        & 79.3 & 92.5 & 93.3 & 97.8 & 0.96 & 73.7 \\
% DPO+CRPO & 56.6 & 82.7 & 86.1 & 94.9 & 0.90 & 48.3 \\

\bottomrule
\end{tabular}
\label{tab:multi-event-ablation}
% \vspace{-20pt}
\end{table}

\section{Results and Analysis}

\subsection{Verifying ALLMs as Judges}

Table~\ref{tab:lalm-instruction-eval} reports ALLM performance on audio understanding benchmarks with ground-truth annotations. 
Recent ALLMs achieve strong performance in both sound event existence and temporal ordering, supporting their suitability as instruction-level evaluators. 
Qwen2.5-Omni-7B performs best overall, followed by Qwen2.5-Omni-3B and Audio Flamingo 3, with consistent gaps across both evaluation aspects.
In terms of benchmark difficulty, ALLMs perform better on AudioTime and the MultiEvent-Temporal benchmarks than on AudioCaps-test and CompA. 
The latter involve real-world audio recordings with overlapping events and background noise, increasing ambiguity in identifying event presence and order. 
In contrast, the more controlled conditions in AudioTime and MultiEvent-Temporal allow more accurate temporal reasoning.
These results indicate that recent ALLMs are suitable as instruction-level evaluators, particularly under controlled or moderately complex conditions. 
Although performance decreases on more ambiguous real-world audio, the judgments remain stable and informative for relative preference construction. 
We therefore use Qwen2.5-Omni-7B as the evaluation-oriented model for instruction-level assessment. 
For preference-based training, we adopt Qwen2.5-Omni-3B as the reward model to avoid the referee-as-player issue while maintaining strong judging capability.

To further validate reliability, we conduct a human verification study on a subset of evaluation samples. 
Human annotators assess sound event existence and temporal ordering using the same protocol. 
We measure event-level agreement and instance-level correlation for sound event existence, and sequence-level agreement for temporal ordering. 
As shown in Table~\ref{tab:human-verification}, ALLM-based judgments show strong consistency with human annotations across all aspects. 
Inter-annotator agreement is also high, with exact-match agreement of 0.92 for sound event existence and 0.95 for temporal ordering.

\subsection{Main Results on Text-to-Audio Generation}

We next evaluate the instruction-following performance of text-to-audio generation models.
Using the ALLM-based evaluation protocol described above, we compare our approach with existing text-to-audio systems across multiple benchmarks.

\noindent\textbf{Results on AudioCaps-test.}
Table~\ref{tab:audiocaps-instruction-eval} reports the evaluation results on AudioCaps-test under the proposed ALLM-based instruction-following protocol.
Overall, our method achieves the strongest performance in following fine-grained textual instructions, particularly in temporal ordering and joint accuracy.
For sound event existence, our model reaches the highest exact match and micro accuracy, achieving performance comparable to TangoFlux-CRPO.
% More importantly, it consistently outperforms strong baselines on temporal order, obtaining the best exact match, pairwise accuracy, and Kendall’s $\tau$.
More importantly, it achieves the best temporal-order exact match and pairwise accuracy, while maintaining a strong Kendall’s $\tau$.
These results indicate that our model more reliably generates multiple target sound events in the correct temporal sequence, rather than matching the text only at a coarse semantic level.
As a result, our method attains the highest joint accuracy, reflecting a stronger overall ability to follow complex instructions that involve both event presence and temporal relationships.

% In terms of common audio quality metrics, our model maintains competitive FAD, FD, IS, and CLAP scores, suggesting that the improvements in instruction following do not come at the cost of degraded audio quality or distributional mismatch.
% Additional predicted aesthetic scores reported on the demo website further support this observation, showing comparable aesthetic quality across systems.
% Overall, these results demonstrate that our approach achieves a better balance between fine-grained instruction adherence and perceptual audio quality.
In terms of standard automatic metrics, our model maintains competitive audio quality and distributional scores, including FAD, FD, KL, and IS, as well as comparable CLAP scores for audio-text alignment. 
This suggests that the improvements in instruction following do not come at the cost of degraded audio quality or semantic alignment. 
% Additional predicted aesthetic~\cite{tjandra2025meta} scores reported on the \href{https://kuan2jiu99.github.io/allm-feedback-tta}{demo website} further support this observation, showing comparable aesthetic quality across systems. 
Additional predicted aesthetic scores provided on the \href{https://kuan2jiu99.github.io/allm-feedback-tta}{project page} further support this observation, showing comparable aesthetic quality across systems.
Overall, these results demonstrate that our approach achieves a better balance between fine-grained instruction adherence and perceptual audio quality.

\noindent\textbf{Results on MultiEvent-Temporal Benchmarks.}
Table~\ref{tab:multi-event-eval} summarizes the results on the MultiEvent-Temporal benchmarks with increasing numbers of target sound events.
As the task becomes more challenging from two to four events, the performance of all models drops noticeably, especially on temporal order and joint accuracy.
Despite this increased difficulty, our method consistently achieves the best results across all settings.
While competing models struggle to preserve correct temporal relationships as the number of events grows, our model maintains substantially higher temporal order accuracy and Kendall’s $\tau$.
This leads to clear gains in joint accuracy, indicating more reliable instruction following under complex multi-event conditions.
Overall, these results show that our approach scales better to longer and more structured temporal instructions, rather than overfitting to simpler multi-event cases.

\noindent\textbf{Results on Sound Scene Story Benchmark.}
We further evaluate all models on the Sound Scene Story benchmark, which features more natural and narrative-style descriptions involving multiple sound events.
Compared to the MultiEvent-Temporal benchmarks, this setting is more realistic and poses additional challenges in both event coverage and temporal coherence.
As shown in Table~\ref{tab:multi-event-eval}, our method achieves the strongest overall performance, outperforming all baselines on temporal order metrics and joint accuracy.
Notably, even strong baseline models exhibit a clear performance gap under this benchmark, highlighting the difficulty of faithfully following narrative instructions.
These results further confirm that our approach generalizes beyond template-based multi-event prompts and is better suited for realistic sound scene generation with complex temporal structure.

\begin{figure}[ht]
    \centering
    \includegraphics[width=0.45\textwidth]{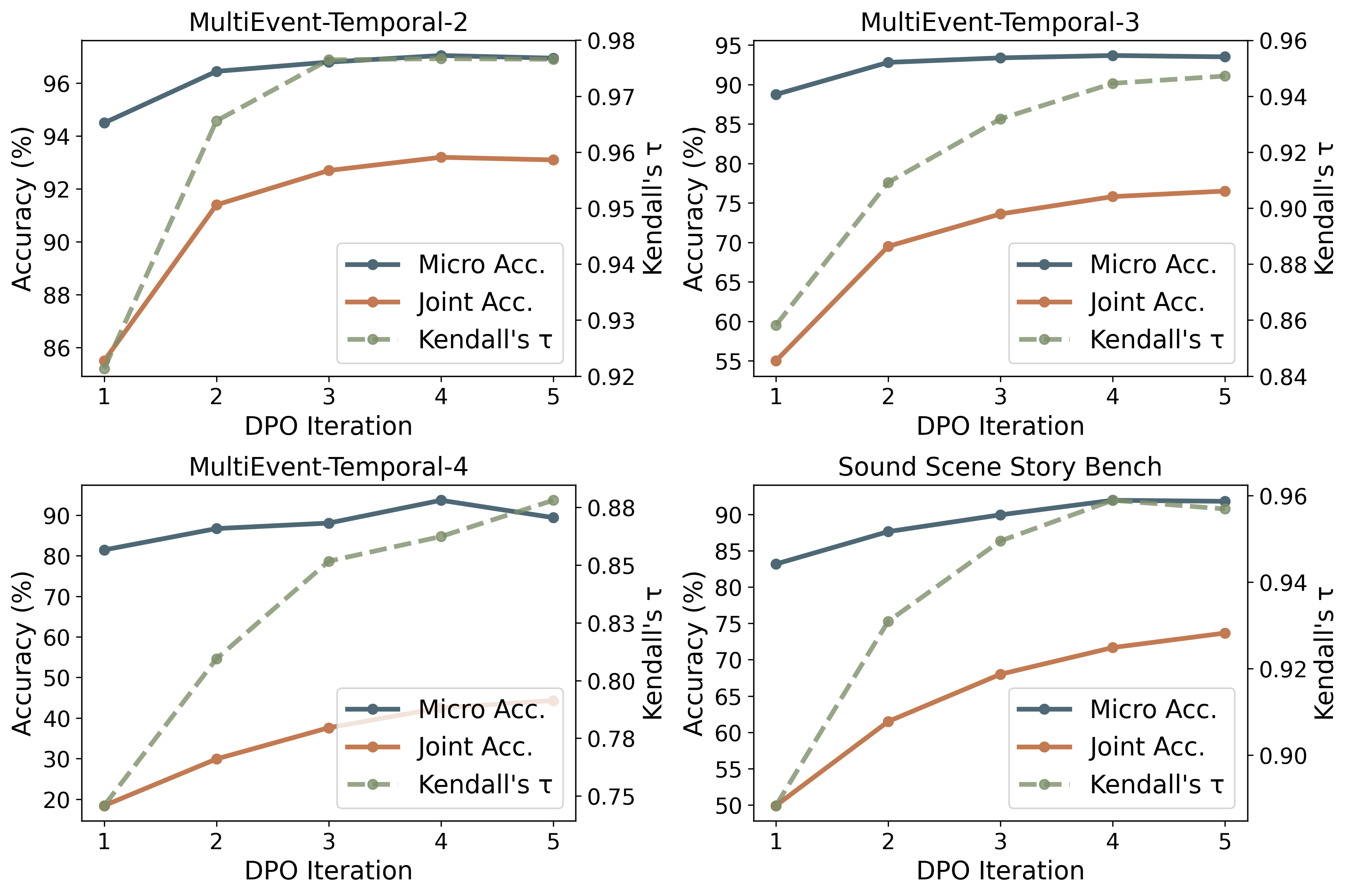} 
    \caption{
    Online DPO progressively improves event existence, temporal reasoning, and overall joint accuracy.
    % across multiple benchmarks. 
    % The improvement is most significant in early iterations and gradually saturates after 3–4 rounds, suggesting diminishing returns from further preference refinement.
    }
    \label{fig:dpo_iteration_exps}
\end{figure}
\subsection{Ablation Study}
For clarity, we denote different preference construction strategies as follows. 
Using ALLM-based fine-grained feedback with direct preference optimization is denoted as ALLM-DPO. 
Replacing the feedback with CLAP-based similarity ranking is denoted as CLAP-DPO. 
Using ALLM-based feedback with supervised fine-tuning is denoted as ALLM-SFT. 
Finally, using the ALLM only for audio captioning followed by caption–text similarity matching is denoted as ALLM-CAP.

% \subsubsection{Is Fine-Grained Feedback Necessary}

% We first examine whether fine-grained ALLM feedback is essential for effective preference construction. 
% To this end, we replace the proposed feedback mechanism with a CLAP-based ranking strategy~\cite{majumder2024tango, hung2024tangofluxsuperfastfaithful}, where generated audio samples are ranked solely by text–audio similarity, and the highest- and lowest-scoring samples are treated as preferred and rejected examples. 
% This setting differs from ALLM-DPO only in how preference pairs are constructed.
% We further design a simplified variant (ALLM-CAP), in which the ALLM is used only for audio captioning. 
% The generated captions are then compared with the ground-truth instruction using a text similarity model like \textit{Qwen3-Embedding-8B}, and samples are ranked accordingly. 
% This variant tests whether decomposed, fine-grained feedback on sound events and temporal relations is truly necessary, or whether caption–text matching alone is sufficient.
% As shown in Tables~\ref{tab:audiocaps-ablation} and~\ref{tab:multi-event-ablation}, ALLM-DPO consistently outperforms both CLAP-DPO and ALLM-CAP in sound event existence and temporal order accuracy. 

% These results indicate that global semantic similarity is insufficient for preference construction. 
% Instead, the ALLM’s ability to explicitly verify target sound events and their temporal relationships plays a critical role in guiding effective optimization.

% \subsubsection{Is Fine-Grained Feedback Necessary}
\subsubsection{Necessity of Fine-Grained Feedback}

We first examine whether fine-grained ALLM feedback is essential for effective preference construction.
To this end, we replace the proposed verifier-based feedback with a CLAP-based ranking strategy~\cite{majumder2024tango, hung2024tangofluxsuperfastfaithful}, where generated samples are ranked solely by global text--audio similarity, and the highest- and lowest-scoring samples are treated as preferred and rejected examples.
This setting differs from ALLM-DPO only in how preference pairs are constructed.
We further design a simplified variant (ALLM-CAP), in which the ALLM is used only for audio captioning.
The generated captions are then compared with the ground-truth instruction using a text similarity model (e.g., \textit{Qwen3-Embedding-8B}), and samples are ranked accordingly.
This variant tests whether decomposed, fine-grained verification of sound events and temporal relations is truly necessary, or whether caption--text matching alone is sufficient.

% As shown in Tables~\ref{tab:audiocaps-ablation} and~\ref{tab:multi-event-ablation}, ALLM-DPO consistently outperforms both CLAP-DPO and ALLM-CAP on sound event existence and temporal order.
As shown in Table~\ref{tab:audiocaps-ablation} and Table~\ref{tab:multi-event-ablation}, ALLM-DPO generally achieves stronger instruction-following performance than CLAP-DPO and ALLM-CAP, with particularly clear gains in joint accuracy and on the more challenging multi-event benchmarks.
% On AudioCaps-test, ALLM-DPO improves the joint correctness score by +2.7 over CLAP-DPO (71.0 vs.\ 68.3), while also achieving higher event existence EM (87.4 vs.\ 86.0).
On multi-event temporal benchmarks, the advantage of fine-grained feedback is amplified:
for MultiEvent-Temporal-3 and -4, ALLM-DPO boosts event existence EM by +14.7 and +10.3 over CLAP-DPO, and yields higher joint scores (e.g., 55.0 vs.\ 37.3 on Temporal-3; 18.4 vs.\ 10.2 on Temporal-4).
These results suggest that global semantic similarity is insufficient for preference construction.

In practice, CLAP-based similarity mainly captures the dominant acoustic concept in an audio clip, but is less sensitive to (i) whether \emph{all} required events are present and (ii) whether their \emph{temporal relation} is correct.
This observation is consistent with prior findings~\cite{ghoshcompa, kuan2026aqascore} that CLAP-style contrastive audio--text representations are effective for coarse semantic alignment, yet can struggle with fine-grained compositional event understanding and temporal reasoning.
As a result, similarity ranking can assign high scores to samples that contain a prominent event but miss other events or swap their order, and this issue becomes more severe as the number of target events increases.
Consequently, CLAP-based similarity does not provide a sufficiently fine-grained reward signal to faithfully reflect instruction-level semantic alignment.
Caption-based ranking provides an intuitive semantic bridge; however, it still relies on a single-sentence summary and a global text-embedding similarity.
Such a coarse bottleneck can easily omit secondary events or blur ordering cues, yielding preferences that are weakly aligned with event-level and relation-level correctness.
The consistently inferior performance of ALLM-CAP further suggests that decomposed verification of sound events and temporal relations is necessary, beyond caption--text matching alone.
% In contrast, the proposed ALLM verifiers explicitly score each target event and its temporal relationship, producing preferences that better reflect the true optimization objective.

% \subsubsection{Comparison with Existing Preference Datasets}

% We next compare ALLM-DPO with two widely used audio preference datasets, Audio-Alpaca and Baton. 
% While models trained on these datasets achieve comparable performance on AudioCaps-test, their performance degrades under more complex scenarios such as MultiEvent-Temporal and S3Bench. 
% In contrast, ALLM-DPO maintains stronger and more stable results, suggesting that fine-grained instruction-level supervision improves generalization to temporally structured tasks.
\subsubsection{Comparison with Existing Preference Datasets}

We next compare ALLM-DPO with two widely used audio preference datasets, Audio-Alpaca~\cite{majumder2024tango} and Baton~\cite{liao2024baton}. 
While models trained on these datasets achieve comparable performance on AudioCaps-test, their performance degrades under more complex scenarios such as MultiEvent-Temporal and S3Bench. 
In contrast, ALLM-DPO maintains stronger and more stable results, suggesting that fine-grained instruction-level supervision improves generalization to temporally structured tasks.
Audio-Alpaca and Baton are static datasets with fixed preference annotations.
In contrast, ALLM-DPO constructs preference pairs dynamically using fine-grained feedback that explicitly verifies target sound events and their temporal relationships. 
This structured supervision allows the model to focus on specific failure modes, such as missing events or incorrect ordering. 
Moreover, the preference data can be tailored to the desired instruction-following objectives, enabling targeted improvement on sound event completeness and temporal correctness.
These results indicate that dynamic, task-aligned preference construction is more effective than relying on static, generic preference datasets for complex instruction-following scenarios.

% \subsubsection{Direct Preference Optimization vs. Supervised Fine-tuning}
\subsubsection{DPO vs. SFT}
We further compare two training objectives within the ALLM-based feedback framework: direct preference optimization (DPO) and supervised fine-tuning (SFT). 
Across benchmarks, ALLM-DPO generally yields larger gains in instruction-following performance. 
In contrast, ALLM-SFT increases FAD and FD on AudioCaps-test, indicating greater deviation from the original data distribution. 
These findings suggest that DPO not only improves instruction adherence but also preserves generation quality more effectively than SFT.

\subsubsection{Online DPO}

Finally, we extend ALLM-DPO to an online setting, where the model trained in one round generates new samples for constructing preference data in the next round.
This process is repeated for five iterations.
As shown in Figure~\ref{fig:dpo_iteration_exps}, online DPO consistently improves sound event completeness (Micro Accuracy), temporal correctness (Kendall’s tau), and overall instruction-level performance (Joint Accuracy), although the gains gradually saturate after three to four iterations.
To examine whether dynamic preference reconstruction is necessary, we compare online DPO with a static variant that reuses the first-round preference pairs for all subsequent iterations.
While the static variant yields initial gains, it deteriorates after the second round: instruction-following accuracy, especially Joint Accuracy, decreases, while FAD rises sharply.
This suggests that fixed preference pairs become stale as the model evolves, causing overfitting to early outputs and degradation in both instruction following and audio fidelity.
In contrast, online DPO regenerates candidates and reconstructs preference pairs at each round, keeping the training signal aligned with the model's current failure modes.
Overall, these results highlight the importance of dynamic preference reconstruction for robust text-to-audio instruction following.

\subsection{Human Evaluation}
For subjective evaluation, we follow a protocol similar to that used in Tango~\cite{ghosal2023tango} and IMPACT~\cite{huangimpact}.
We randomly select 50 generated audio samples from AudioCaps-test and S3Bench, covering both basic and advanced instruction settings.
Each sample is evaluated along two dimensions: text relevance (REL) and overall quality (OVL), using a 1-to-5 rating scale.
All samples are rated by three human participants.
We include three systems in the human evaluation: the TangoFlux base model, our proposed ALLM-DPO method, and a CLAP-DPO variant.
This comparison aims to examine whether our approach yields improvements not only on ALLM-based and other automatic metrics, but also from a human perspective.
The results are summarized in Table~\ref{tab:audiocaps-ablation}.
Overall, ALLM-DPO achieves higher REL (4.28 vs. 3.55 for CLAP-DPO) and slightly higher OVL scores in human evaluation. 
% slightly higher OVL scores (3.39 vs. 3.32 for CLAP-DPO) in human evaluation.
The gain in text relevance does not come at the expense of overall audio quality. 
These results indicate that our method improves not only automatic metrics but also human-perceived text relevance and overall quality.

% \subsection{Online Direct Preference Optimization}

% \subsection{Improvement and Regression Analysis}

% Regression is measured only on base-perfect instances, where the base model is fully correct (all events identified or Kendall’s Tau equals 1), and is defined as cases where the new model fails.
% Improvement is measured only on base-imperfect instances and is defined as cases where the new model corrects at least one error.
% This design ensures that regression and improvement reflect stability and error correction, respectively, without being diluted by instances where improvement is impossible.

% Error analysis
% Human Evaluation (?) -> win rate?

\section{Conclusion}

% In this work, we study a fundamental limitation of current TTA generation systems: despite strong perceptual quality, they often fail to correctly follow fine-grained instructions involving multiple sound events and temporal relationships.
% To address this gap, we propose a fine-grained evaluation and training framework that leverages large audio-language models as judges.
% By assessing sound event existence and temporal order, ALLMs provide explicit, instruction-aware feedback that can be directly used to construct preference data and guide training via direct preference optimization.
% Experiments show consistent improvements in sound event completeness and temporal ordering across both existing benchmarks and the proposed Sound Scene Story Benchmark.
% Our results suggest that instruction-aware feedback offers a scalable and effective way to improve controllability in text-to-audio generation.
% Beyond perceptual realism, this work highlights the importance of evaluating and training TTA systems based on whether they truly follow user instructions, opening new directions for more reliable and controllable audio generation.

We revisit TTA generation from the perspective of instruction-level correctness. 
While recent systems achieve strong perceptual realism, our study shows that they often overlook fine-grained requirements such as sound event completeness and temporal ordering. 
By leveraging audio-aware LLMs as structured judges, we demonstrate that instruction-aware feedback can serve as an effective training signal for improving controllability. 
Our results suggest that moving beyond global similarity toward explicit verification of target events and their relations is crucial for reliable text-to-audio generation.
More broadly, this work highlights the importance of aligning evaluation and training objectives with user intent, pointing toward scalable, model-based feedback mechanisms for more controllable and interactive audio generation.

\section{Disclosure of Generative AI Use}
% AI-assisted tools were used solely to improve the clarity and fluency of the manuscript.
The authors used Claude and ChatGPT to assist with grammar checking, language polishing, and improving the readability of the manuscript. 

\bibliographystyle{IEEEtran}
\bibliography{mybib}

\end{document}